\documentclass[aps,pra,twocolumn,superscriptaddress]{revtex4}

\usepackage{color}
\usepackage{bm}
\usepackage{rotating}
\usepackage{array,multirow,graphicx}
\usepackage{tabularx}

\newcolumntype{M}[1]{>{\centering\arraybackslash}m{#1}}

\usepackage{algorithm2e}
\usepackage{newtxmath,newtxtext}
\DeclareMathAlphabet{\mathcal}{OMS}{cmsy}{m}{n}

\usepackage{soul}
\usepackage{mathtools}
\newcommand{\mathdash}{\relbar\mkern-9mu\relbar}
\DeclareMathAlphabet\mathbfcal{OMS}{cmsy}{b}{n}

\newcommand{\ra}{\rangle}

\begin{document}

\title{Variational quantum metrology 
for multiparameter estimation under dephasing noise}

\author{Trung Kien Le}
\affiliation{Department of Physics, University of California, Santa Barbara, USA}
\author{Hung Q. Nguyen}
\affiliation{Nano and Energy Center, VNU University of Science, Vietnam National University, Hanoi}
\author{Le Bin Ho}
\email{binho@fris.tohoku.ac.jp}
\affiliation{Frontier Research Institute for Interdisciplinary Sciences, 
Tohoku University, Sendai 980-8578, Japan}
\affiliation{Department of Applied Physics, 
Graduate School of Engineering, 
Tohoku University, Sendai 980-8579, Japan}
\
% \date{\today}%
\begin{abstract}
We present a hybrid quantum-classical variational 
scheme 
to enhance precision in quantum metrology. 
In the scheme, both the initial state 
and the measurement basis 
in the quantum part are parameterized 
and optimized via the classical part. 
It enables the maximization of information gained 
about the measured quantity. 
We discuss specific applications to 
3D magnetic field sensing 
under several dephasing noise modes. 
Indeed, we demonstrate its ability 
to simultaneously estimate all parameters 
and surpass the standard quantum limit, 
making it a powerful tool for metrological applications.
\end{abstract}
\maketitle

\section{Introduction}
Quantum metrology is an estimation process 
that utilizes unique quantum phenomena 
such as entanglement and squeezing 
to improve the precision of estimation 
beyond classical limits
\cite{PhysRevLett.72.3439,
Giovannetti2011,Toth_2014}.
Recent development in quantum computing 
leads to numerous optimal algorithms 
for enhancing precision in 
single-parameter estimation,
such as adaptive measurements
\cite{Barndorff-Nielsen_2000,Fujiwara_2006,Zhang_2018,
PhysRevX.7.041009%,
%https://doi.org/10.1002/qute.202100080
}, 
quantum error correction
\cite{%https://doi.org/10.1002/qute.202100080,
PhysRevLett.112.150802,Zhou2018}, 
and optimal quantum control
\cite{%https://doi.org/10.1002/qute.202100080,
PhysRevLett.128.160505,Pang2017,Yang2021}. 
So far, a variational algorithm has been demonstrated 
by combining the advantages of both quantum 
and classical systems for quantum-enhanced metrology
\cite{Koczor_2020,9605341,Yang2021,
PhysRevLett.123.260505}.
A similar protocol for spin systems was also
introduced \cite{PhysRevX.11.041045,Zheng2022}.

Multiparameter estimation is 
essential in various fields, 
such as Hamiltonian tomography \cite{PhysRevLett.124.160502}, 
multiphase sensing \cite{PhysRevLett.116.030801,
PhysRevA.102.022602,
PhysRevLett.125.020501}, 
gravitational wave detection \cite{Schnabel2010}, 
and atomic clocks \cite{RevModPhys.83.331,
RevModPhys.87.637}. 
However, the estimation is more challenging 
due to the incompatibility \cite{PhysRevX.12.011039}. 
In these cases, simultaneous 
determination of all parameters 
is impossible, resulting in a tradeoff \cite{Kull_2020}. 
Numerous techniques have been 
introduced to tackle this challenge, 
including establishing 
optimal measurement strategies \cite{PhysRevLett.119.130504,PhysRevA.100.032104}, 
employing parallel scheme \cite{PhysRevLett.125.020501},
sequential feedback scheme \cite{PhysRevLett.117.160801}, 
and implementing 
post-selection procedures \cite{10.1063/5.0024555}.
More recently, a variational toolbox 
for multiparameter estimation was proposed
\cite{Meyer2021}, which is a 
generalization from the previous work mentioned above
\cite{9605341}. 

%
%It has a wide range of applications,
%from quantum magnetometry
%\cite{PhysRevA.102.022602,PhysRevLett.125.020501},
%to quantum clocks \cite{RevModPhys.83.331,RevModPhys.87.637},
%quantum imaging \cite{Lugiato_2002,Moreau2019},
%gravitational-wave detection
%\cite{Schnabel2010}, 
%dark matter detection 
%\cite{10.1007/978-3-030-31593-1_5},
%and so on.
%So far, it was demonstrated %the 
%quantum-enhanced beyond the Heisenberg limit
%with nonlinear interaction 
%\cite{PhysRevLett.102.100401,PhysRevX.2.041006,PhysRevA.89.022107},
%interaction-based \cite{Napolitano2011,PhysRevLett.119.193601},
%and even without entanglement \cite{RevModPhys.90.035006}.
%%The studies of 
%Quantum metrology 
%under noisy environments
%\cite{PhysRevLett.112.120405,Tsang_2013,Haa},
%post-selection measurements
%\cite{PhysRevLett.114.210801,
%Arvidsson-Shukur2020,doi:10.1063/5.0024555}, 
%and quantum error correction
%\cite{PhysRevLett.112.150802,
%Gorecki2020optimalprobeserror,
%PhysRevResearch.2.013235}
%are also extensively reported.
%

While using variational schemes is promising, 
their potential significance 
in multiparameter quantum metrology 
has yet be fully understood, even in principle. 
Furthermore, determining the optimal quantum resources 
and measurement strategy to extract maximum information 
about all parameters is limited by the tradeoffs 
in estimating incompatible observables 
\cite{Kull_2020,Zhu2015,PhysRevA.94.052108}
and required collective measurements over 
multiple copies of a probe state
\cite{Zhu2015,PhysRevA.94.052108}.
Therefore, finding a suitable and practical strategy 
for precise estimation of multiple parameters remains a
thriving area of quantum metrology.

%One of the current challenges in quantum metrology 
%is the simultaneous accurate estimation of multiphases 
%using a single quantum system. This is due to the incompatability of optimal local observables, a well-known problem in multiparameter quantum metrology. 

%To overcome this problem, 
%various approaches have been proposed, including semidefinite programming with linear constraints \cite{PhysRevX.11.011028}, variational ansatz for classical Fisher information, 

%In this work, we propose a hybrid classical-quantum algorithm together with quantum metrological bounds as cost functions.  Our bounds simultaneously minimize the classical and quantum Fisher information, as well as ensure the attainability of quantum bounds of our ansatz following classical training. 

In this work, we propose a variational scheme
to enhance the precision of 
multiparameter estimation 
in the presence of dephasing noise. 
The basic idea is to use 
a quantum computer to prepare
a trial state (an ansatz) that depends on 
a set of trainable variables. 
The state is subjected to 
a series of control operations, 
representing unknown multiparameter and noise, 
and then is measured through observables 
determined by other trainable variables. 
The measurement results are used to update 
the trainable variables and optimize 
the estimation of the unknown parameters.
%The state is then evolved under 
%a set of control operations, 
%which imprints 
%a set of unknown parameters
%and noise.
%The final state is then
%measured using a set of observables, 
%also parameterized by
%classical variables. 
%The measurement results are then used
%to update the classical variables 
%to optimize the estimation
%of the unknown multiparameter. 

Optimizing both the initial probe state  
and the measurement operators allows us to 
identify suitable conditions for the quantum probe 
to increase sensitivity and achieve 
the ultimate quantum limit for all parameters.
In numerical simulations, 
we estimate a 3D magnetic field under 
a dephasing noise model 
and find that sensitivity for all parameters 
can simultaneously reach the ultimate quantum bound,
i.e., the classical bound equals the quantum bound.
We also examine a time-dependent 
Ornstein-Uhlenbeck model \cite{PhysRev.36.823}
and observe results surpassing 
the standard quantum limit by 
increasing the probe's number of particles. 
This approach holds promise 
for a wide range of metrological applications, 
including external field sensing, 
precision spectroscopy, 
gravitational wave detection, 
and others, where the effects of 
noises cannot be ignored.

\section{Results}
\subsection{Variational quantum metrology}
The goal of multiparameter estimation
is to evaluate 
a set of unknown $d$ parameters
$\bm \phi = (\phi_1, \phi_2, 
\cdots, \phi_d)^\intercal$,
%These parameters 
which are imprinted 
onto a quantum probe 
via a unitary evolution
$\bm U(\bm\phi) = \exp(-it\bm H\bm\phi)
= \exp(-it\sum_{k=1}^dH_k\phi_k)$,
where $\bm H = (H_1, H_2, \cdots, H_d)$
are non-commuting Hermitian Hamiltonians.
The precision of 
estimated parameters $\Check{\bm\phi}$ 
is evaluated using 
a mean square error matrix (MSEM)
$
%\begin{align}
     V%(\bm \Pi,\check{\bm\phi})
    = \sum_m p(m|\bm\phi)
    \big[\check{\bm\phi}(m)-\bm\phi\big]
    \big[\check{\bm\phi}(m)-\bm\phi\big]^\intercal,
%\end{align}
$
where $p(m|\bm\phi) = {\rm Tr}[\rho(\bm\phi) E_m]$ 
is the probability
for obtaining an outcome $m$
when measuring the 
final state $\rho(\bm\phi)$
by an element $E_m$ in a positive,
operator-value measure (POVM).
For unbiased estimators,
the MSEM %becomes a covariance matrix 
%which 
obeys the %classical 
Cram{\'e}r-Rao bounds (CRBs)
\cite{helstrom1976quantum,
holevo2011probabilistic, conlon2022gap,hayashi2023tight}
\begin{align}\label{eq:RB}
    {\rm Tr}\big[WV\big]%(\bm \Pi,\check{\bm\phi})
    \ge \mathsf{C_F}
    \ge \mathsf{C_{NH}}
    \ge \mathsf{C_H}
    \ge %\max(\mathsf{C_S^S,C_S^R)},
    \mathsf{C_S},
\end{align}
where $W$ is a scalar weight matrix,
which can be chosen as 
an identity matrix without loss of generality.
The classical bound is $\mathsf{C_F} 
= {\rm Tr}[
{WF}^{-1}]$,
where $F$ is the classical 
Fisher information matrix (CFIM)
with elements $F_{ij} 
= \sum_m 
\frac{1}{p(m|\bm\phi)}
[\partial_{\phi_i} p(m|\bm\phi)]
[\partial_{\phi_j} p(m|\bm \phi)]$
\cite{paris2009quantum}.
The Nagaoka–Hayashi bound 
$\mathsf{C_{NH}}$ and
Holevo bound $\mathsf{C_H}$
are given via
semidefinite programming, %(SPD), 
i.e., 
$\mathsf{C_{NH}} = 
\underset{\{D, X\}}{\min}
%\big(
{\rm Tr}[(W\otimes\rho(\bm\phi))\ \cdot D]
%\big)
$
\cite{Conlon2021,hayashi2023tight}, 
and 
$\mathsf{C_H} = 
\underset{\{X\}}{\min}
\big(
{\rm Tr}[W{\rm Re}Z+
\vert \vert\sqrt{W}
{\rm Im}Z\sqrt{W}]\vert \vert_1
\big)$
\cite{holevo2011probabilistic,hayashi2005},
where $D$ is a $d$-by-$d$ matrix contains Hermitian
operators $D_{ij}$ and satisfies 
$D \ge XX^\intercal$,
$X = (X_1, X_2,\cdots,X_d)^\intercal$
satisfies ${\rm Tr}[\rho(\bm\phi)X_j] = \phi_j$
and 
${\rm Tr}[X_i\partial_{\phi_j}\rho(\bm\phi)]
=\delta_{ij}$, 
$Z$ is a positive semidefinite matrix
with elements $Z_{ij} = {\rm Tr}[X_iX_j\rho(\bm\phi)]$.
Finally, $\mathsf{C_S} 
= {\rm Tr}[WQ^{-1}]$
%and 
%$\mathsf{C_S^R} 
%= {\rm Tr}[W{\rm Re}(Q^\mathsf{R})^{-1}]
%+\vert \vert\sqrt{W}{\rm Im}(Q^\mathsf{R})^{-1}
%\sqrt{W}\vert \vert_1
%$
is a symmetric logarithmic derivative 
(SLD) quantum bound where
$Q_{ij} = {\rm Re}
\big[{\rm Tr}[\rho(\bm\phi)L_iL_j]\big]$
is the real symmetric quantum 
Fisher information matrix (QFIM)
that defined through the SLD
$2\partial_{\phi_j}\rho(\bm\phi)
= \{L_j,\rho(\bm\phi)\}$
\cite{paris2009quantum}.
%and 
%$Q^\mathsf{R}_{ij} = 
%{\rm Tr}[\rho(\bm\phi)\tilde{L}_i
%\tilde{L}_j^\dagger]$
%is defined by the right 
%logarithmic derivative
%$\partial_j\rho 
%= \rho\tilde{L}_j$
%\cite{1055103}.

Although optimal estimators can achieve 
$\mathsf{C_F}$ \cite{kay1993estimation},
the $\mathsf{C_{NH}}$ 
can be attainted with separable measurements
for qubits probes \cite{Conlon2021},
and asymptotic achievement of 
$\mathsf{C_H}$ is possible 
\cite{Yang2019, 10.1214/13-AOS1147,PhysRevX.11.011028,
https://doi.org/10.48550/arxiv.2008.01502,PhysRevLett.123.200503}, 
%it is impossible to attain 
it is not always possible to attain
$\mathsf{C_S}$ for 
multiparameter estimation \cite{hayashi2005}.
In this work, we attempt to reach this bound.
In some instances, $\mathsf{C_H = C_S}$ 
if a weak commutativity condition 
${\rm Im(Tr}[L_jL_i\rho(\bm\phi)]) = 0$ 
is met \cite{hayashi2005, 
Demkowicz-Dobrzanski_2020}. 
A similar
condition 
for pure states
is also applied
to attain $\mathsf{C_F = C_S}$ 
\cite{K_Matsumoto_2002,PhysRevLett.111.070403,PhysRevLett.116.030801}. 
Further discussion on
the interplay between 
$\mathsf{C_{NH}}, \mathsf{C_H},$
and $\mathsf{C_S}$ 
has been reported \cite{conlon2022gap}.
However, this condition alone 
is insufficient to achieve 
the quantum bounds practically; 
instead, attaining 
 $\mathsf{C_S}$
and $\mathsf{C_H}$ 
also requires 
entangled measurements (POVM)
over multiple copies
\cite{PhysRevX.11.011028,
https://doi.org/10.48550/arxiv.2008.01502}.
Recently, Yang et al., have derived 
saturation conditions for general POVMs
\cite{PhysRevA.100.032104}.
To be more precise, when 
$F \ge Q > 0$ and for any arbitrary 
full-rank positive weight matrix $W > 0$, 
the equally $\mathsf{C_F = C_S}$ implies $F = Q$.

\begin{figure}[t!]
\centering
\includegraphics[width=8.6cm]{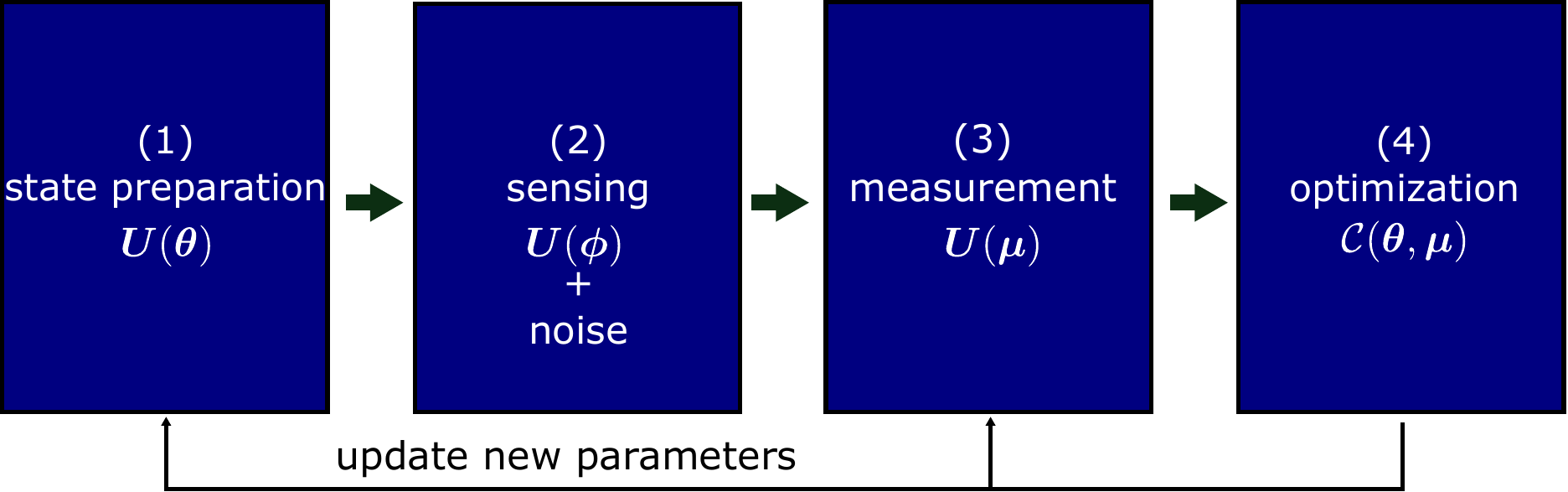}
\caption{
\textbf{Variational quantum metrology}.
(1) use quantum circuit $\bm U(\bm\theta)$ 
to prepare a variational state;
(2) encode multiparameter $\bm\phi$ 
and noise using $\bm U(\bm\phi)$ 
and noise channels;
(3) use circuit $\bm U(\bm\mu)$ 
to create a variational POVM 
for measurement;
(4) send measurement results 
to a classical computer to 
optimize cost function 
$\mathcal{C}(\bm\theta,\bm\mu)$ 
using a gradient-based optimizer. 
Update new training variables and repeat 
the scheme until it converges.}
\label{fig:1}
\end{figure} 

This paper presents a variational 
quantum metrology (VQM) scheme 
following Meyer et al. toolbox \cite{Meyer2021}
as sketched in Fig.~\ref{fig:1} to
optimize both the preparation state 
and POVM. % to attend the ultimate 
%quantum bound, i.e., $\min(\mathsf{C_S^S,C_S^R)}$.
A quantum circuit $\bm U(\bm\theta)$
is used to generate a 
variational preparation state 
with trainable variables $\bm\theta$.
Similar quantum circuit with 
variables $\bm \mu$ 
is used to generate a variational POVM
$\bm E(\bm\mu) 
= \{E_m(\bm\mu) 
= \bm U^\dagger(\bm\mu)E_m \bm U(\bm\mu) > 0 
\big| \sum_m E_m(\bm\mu) = \bm I\}$.
Using classical computers, 
a cost function $\mathcal{C}(\bm\theta,\bm\mu)$
can be optimized %to generate 
to update the 
variables for quantum circuits, 
resulting in enhanced information extraction.
The scheme is repeated until it converges.

\begin{figure}[t]
\centering
\includegraphics[width=8.6cm]{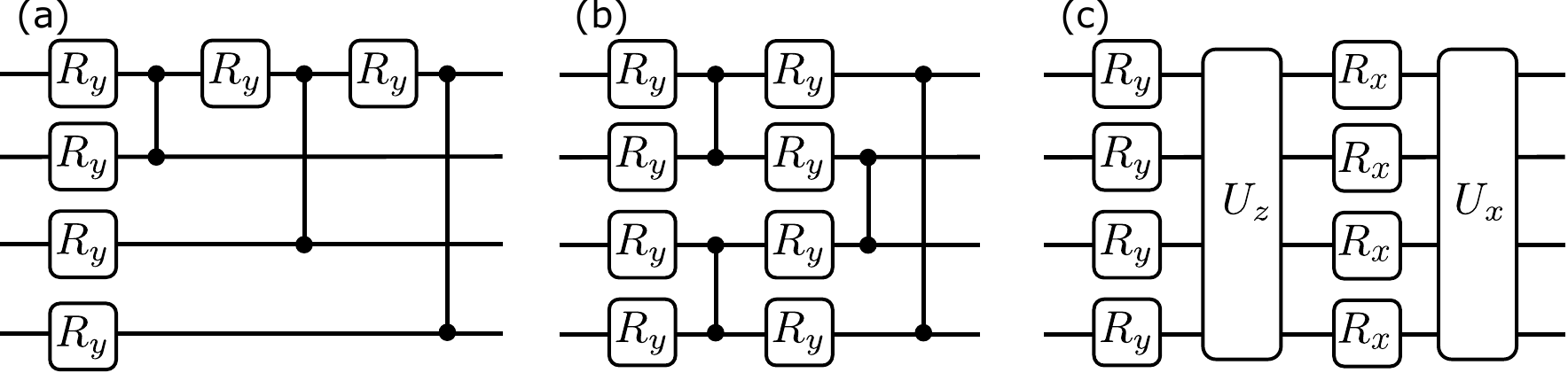}
\caption{
\textbf{Ansatzes for preparation state and POVM}.
(a) star topology entangled ansatz.
(b) ring topology entangled ansatz.
(c) squeezing ansatz.
In the circuits, $R_{x(y)}$: $x(y)$-rotation gate, 
$U_{x(z)}$: global M{\o}lmer--S{\o}rensen gate,
$\bullet\!\!\!\mathdash\!\!\!\bullet$: controlled-Z gate.
}
\label{fig:2}
\end{figure} 
%

%\begin{widetext}
\begin{figure*}[t]
\centering
\includegraphics[width=16.0cm]{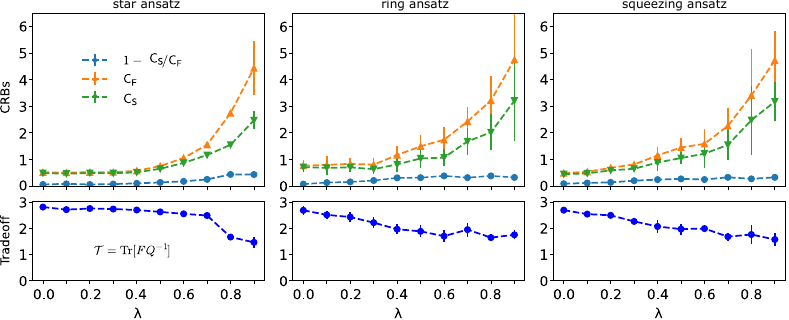}
\caption{
\textbf{Variational quantum metrology under dephasing noise}.
%The star and ring ansatzes were kept at two layers, and the squeezing ansatz was kept at one layer. 
(Top): plot of the optimal cost function
$\mathcal{C} (\bm \theta, \bm\mu)$,
classical bound $\mathsf{C_F}$, 
and SLD quantum bound $\mathsf{C_S}$
as functions of dephasing probability.
From left to right: star, ring,
and squeezing ansatz.
(Bottom): plot of 
corresponding tradeoff 
$\mathcal{T}$.
Numerical results are calculated
at $N = 3$, the optimal number of layers 
in Fig.~\ref{fig:8},
and the results are 
averaged after 10 samples.
}
\label{fig:3}
\end{figure*} 

To investigate the attainable 
the ultimate SLD quantum bound, 
we define the cost function 
by a relative difference
\cite{PhysRevLett.123.200503}
\begin{align}\label{eq:cost}
\mathcal{C} (\bm \theta, \bm\mu) = 
1 - \dfrac{\mathsf{C_S}}
{\mathsf{C_F}},
\end{align}
which is positive semidefinite 
according to Eq.~\eqref{eq:RB}.
The variables
are trained by
solving the optimization task
$\underset{\{\bm\theta, \bm\mu\}}{\arg\min}\
\mathcal{C} (\bm\theta, \bm\mu)$.
As the value of $\mathcal{C}(\bm \theta, \bm\mu)$ 
approaches zero, we reach the ultimate SLD
quantum bound where $\mathsf{C_F}= \mathsf{C_S}$.
Notably, we strive for 
agreement between classical 
and SLD quantum bounds assuming 
${\rm Tr}[WV] = \mathsf{C_F}$, 
and thus, omit discussion on 
the estimator for achieving 
${\rm Tr}[WV] = \mathsf{C_F}$.
Moreover, the cost function \eqref{eq:cost}
serves as a technical tool to 
optimize the variational scheme,
while the main analyzing quantities are CRBs.
A vital feature of the VQM is 
using variational quantum circuits, 
which allows for optimizing the
entangled probe state and 
measurements
to extract the maximum information 
about the estimated parameters.
This approach thus does not require 
entangled measurements over multiple copies.
We further discuss various cost functions
in the Discussion section.

\subsection{Ansatzes}
We propose three variational circuits:
a star topology ansatz, 
a ring topology ansatz, and a squeezing ansatz. 
The first two ansatzes are inspired by 
quantum graph states, 
which are useful resources for quantum metrology
\cite{PhysRevLett.124.110502,PhysRevA.102.052601}.
A conventional graph state is formed 
by a collection of vertices $V$ 
and edges $D$ as $G(V,D) = 
\prod_{{i,j} \in D} {\rm CZ}^{ij} 
|+\rangle ^{V}$, where ${\rm CZ}^{ij}$ 
represents the controlled-Z gate 
connecting the $i$ and $j$ qubits, 
and $|+\rangle$ is an element 
in the basis of Pauli $\sigma_x$.
The proposed ansatzes here incorporate 
$y$-rotation gates 
($R_y(\theta) = e^{-i\theta\sigma_y/2}$)
at every vertex prior to CZ gates
(see Fig.~\ref{fig:2}a,b).
The squeezing ansatz 
in Fig.~\ref{fig:2}c is inspired by 
squeezing states, which is another 
useful resource for quantum metrology
\cite{Maccone2020squeezingmetrology,
Gessner2020,PhysRevA.102.062610}. 
It has $x(y)$-rotation gates 
and global M{\o}lmer--S{\o}rensen 
gates $U_{x(z)}$, where
$U_{x(z)} = \exp(-i\sum_{j=1}^N\sum_{k = j+1}^N
\sigma_{x(z)}\otimes \sigma_{x(z)}
\frac{\chi_{jk}}{2})$ 
for an $N$-qubit circuit
\cite{PhysRevLett.82.1835}.
The trainable variables for one layer are 
$2N - 2$, $2N$, and $N(N+1)$ 
for the star, ring, 
and squeezing ansatz, respectively.
Hereafter, we use these ansatzes for generating 
variational preparation states
and variational POVM in the VQM scheme.

\subsection{Multiparameter estimation under dephasing noise}
After preparing a variational state
$\rho(\bm\theta) = \bm U(\bm\theta)
\rho_0\bm U(\bm\theta)$,
we use it to estimate a 3D 
magnetic field under dephasing noise. 
The field is imprinted onto every single qubit via 
the Hamiltonian $\bm H = \sum_{i \in\{x,y,z\}}\phi_i\sigma_i$,
where $\bm \phi = (\phi_x, \phi_y, \phi_z)$,
and $\sigma_i$ is a Pauli matrix.
Under dephasing noise, 
the variational state 
$\rho (\bm\theta)$ evolves to
\cite{Koczor2020}
\begin{align}\label{eq:ev}
    \mathcal{E}_t(\rho) = 
    \Bigg[\prod_{k=1}^N 
    e^{\gamma t\mathcal{L}^{(k)}}\Bigg]
    e^{-it\mathcal{H}}\rho,
\end{align}
where we omitted $\bm\theta$ 
in $\rho(\bm\theta)$ for short.
The superoperator $\mathcal{H}$ 
generates a unitary dynamic 
$\mathcal{H}\rho = [\bm H, \rho]$,
and $\mathcal{L}^{(k)}$ is a
non-unitary dephasing superoperator
with $\gamma$ is the decay rate.
In terms of Kraus operators, 
the dephasing superoperator gives
\begin{align}\label{eq:eLKrau}
    e^{\gamma t\mathcal{L}^{(k)}}\rho
    = K_1^{(k)}\rho [K_1^{(k)}]^\dagger 
    + K_2^{(k)}\rho [K_2^{(k)}]^\dagger,
%    & = \dfrac{1}{2}(1+\sqrt{1-\gamma})\rho 
%    + \dfrac{1}{2}(1-\sqrt{1-\gamma})\sigma_z\rho\sigma_z 
\end{align}
where $K_1 = \begin{pmatrix} 
\sqrt{1-\lambda} & 0\\0&1\end{pmatrix}$
and $K_2 = \begin{pmatrix} 
\sqrt{\lambda} & 0\\0&0\end{pmatrix}$
are Kraus operators,
and $\lambda = 1-e^{-\gamma t}$
is the dephasing probability.
Finally, the state is measured 
in the variational POVM $\bm E(\bm\mu)$
and yields the probability
$p(m) = {\rm Tr}[\mathcal{E}_t(\rho)E_m(\bm\mu)]$.
Note that $p$ also depends on $\bm\theta,
\bm\phi$, and $\bm\mu$.

It is important to attain 
the ultimate SLD quantum bound,
i.e., $\mathsf{C_F} = \mathsf{C_S}$.
We thus compare numerical 
results for the cost function, 
$\mathsf{C_F}$, and 
$\mathsf{C_S}$ 
as shown in the top panels of
Fig.~\ref{fig:3}.
At each $\lambda$,
the cost function and 
other quantities are plotted 
with the optimal $\theta$ obtained
after stopping the training by EarlyStopping
callback \cite{keras_early_stopping}.
The numerical results are presented
at $N = 3$, and the number of layers
is chosen from their optimal values
as shown in the Method
and Fig.~\ref{fig:8}.
Through the paper, 
we fixed $(\phi_x, \phi_y, \phi_z) = 
(\pi/6, \pi/6, \pi/6)$.

We find that for small noises, 
$\mathsf{C_F}$ reaches %the 
%ultimate SLD quantum bound,
$\mathsf{C_S}$,
which is consistent with 
earlier numerical findings 
\cite{https://doi.org/10.48550/arxiv.2008.01502,
PhysRevLett.123.200503}.
Remarkably, different from the previous findings 
where the convergence of these bounds is not clear,
here we show that both $\mathsf{C_F}$
and $\mathsf{C_S}$ 
remain small 
(also in comparison to previous work \cite{Meyer2021}) 
without any divergence.
We further compare the performance 
of the star ansatz to that of 
the ring and squeezing ansatzes.
%The star ansatz 
It saturates 
the ultimate quantum limit 
for dephasing probabilities 
$\lambda < 0.5$, 
whereas the ring and 
squeezing ansatzes only 
reach the limit for 
$\lambda < 0.2$. 
The reason is that the star graph exhibits a central vertex 
connected to the remaining $N-1$ 
surrounding vertices, which facilitates 
robust quantum metrology, 
as discussed in \cite{PhysRevLett.124.110502}.

Furthermore, we evaluate the 
tradeoff between the CFIM and QFIM 
by introducing a function
$
%\begin{align}
    \mathcal{T} = \textrm{Tr}[FQ^{-1}].
%\end{align}
$
For unknown $d$ parameters, 
the naive bound is $\max (\mathcal{T}) = d$,
leading to simultaneous optimization 
of all parameters. %with $F$ and $Q$ being equal.
The results are shown in 
the bottom panels of Fig. \ref{fig:3}
and agree well with 
the CRBs presented in the top panels,
wherein $\mathcal{T}\to 3$ whenever 
the SLD quantum bound is reached. 
So far, 
we observe that $\mathcal{T}> d/2$ 
for all cases, which is 
better than the theoretical prediction
previously
\cite{Vidrighin2014}. 
This observation exhibits a practical advantage 
of the VQM approach across 
different levels of noise.

\begin{figure}[t]
\centering
\includegraphics[width=8.6cm]{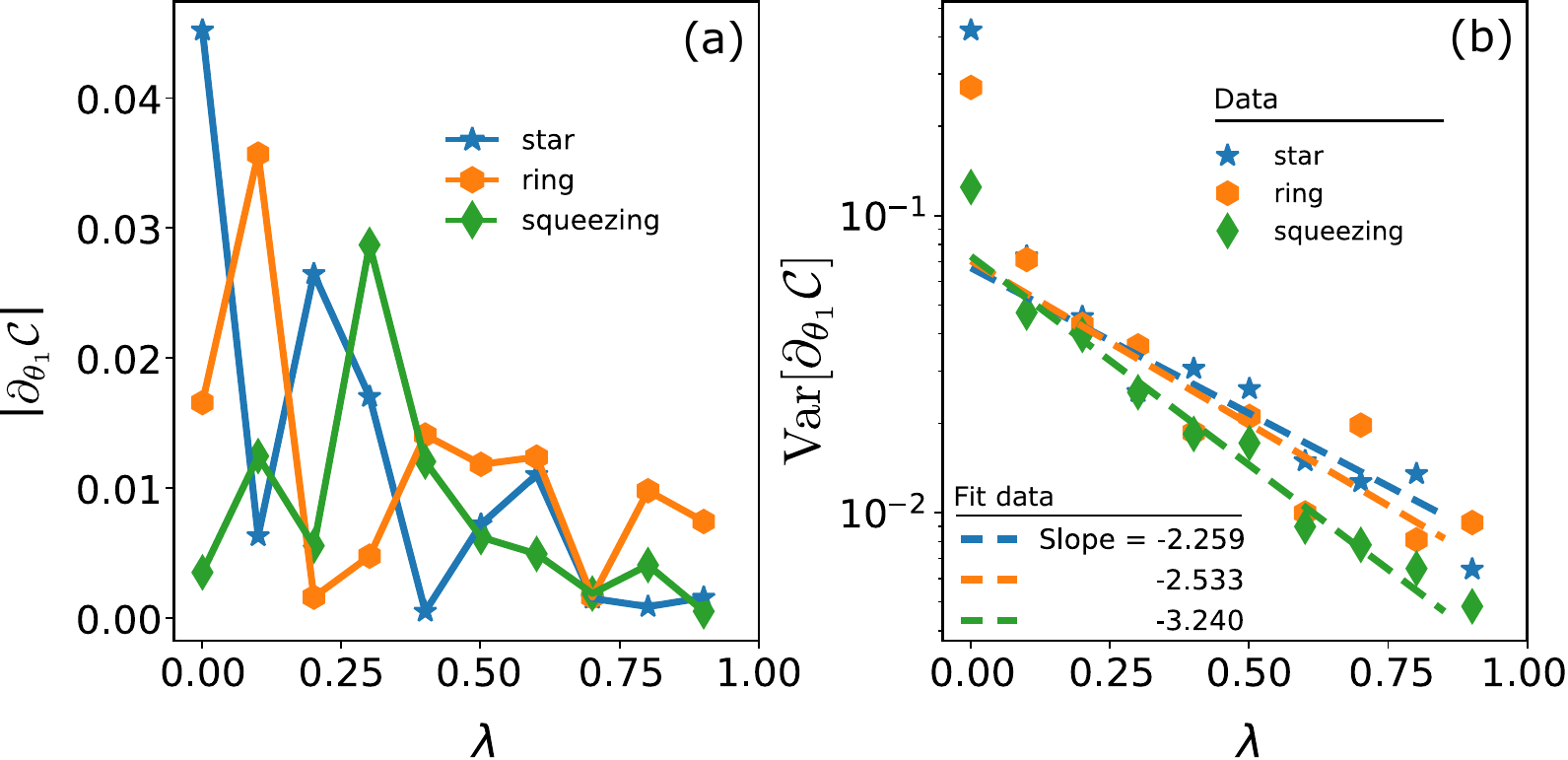}
\caption{
\textbf{Barren plateau}. 
(a) Plot of  
$|\partial_{\theta_1}\mathcal{C}|$ 
as a function of 
the dephasing probability $\lambda$.
(b) Plot of the variance of gradient 
Var[$\partial_{\theta_1}\mathcal{C}$].
%is plotted as a function of 
%the dephasing probability $\lambda$. 
The slope of each fit line indicates 
the exponential decay of the gradient,
which is a sign of the barren plateau effect.
The results are taken average after 200 runs.
}
\label{fig:4}
\end{figure}

\subsection{Barren Plateaus}
Variational quantum circuits under 
the influence of noises 
will exhibit a barren plateau (BP), 
where the gradient along any direction of 
the variables' space vanishes exponentially 
with respect to 
%the number of qubits or 
the noise level
\cite{Wang2021}. The BP prevents reaching
the global optimization of the training space,
thereby reducing the efficiency 
and trainability of 
the variational quantum circuit.
However, BPs can be 
partially mitigated through carefully 
designing ansatzes and 
cost functions \cite{Cerezo2021}.

The deviation of 
CRBs shown in Fig.~\ref{fig:3} 
may be subject to the BP raised by noise.
We examine such dependent and show
the results in Fig. \ref{fig:4}.
We plot $|\partial_{\theta_1} \mathcal{C}|$
(Fig. \ref{fig:4}a) and 
Var$[\partial_{\theta_1} 
\mathcal{C}]$ (Fig. \ref{fig:4}b),
where $\mathcal{C}$ is defined in 
Eq. \eqref{eq:cost}
%To estimate Var$[\partial_{\theta_1} 
%\mathcal{C}]$, we perform 
after 200 runs with random initialization 
of $\bm \theta$ and $\bm \mu$
for each value of $\lambda$. 
As predicted, both of them demonstrate an exponential 
decline with an increase in the dephasing probability.
Especially, Var$[\partial_{\theta_1} 
\mathcal{C}]$ exponentially vanishes 
with the slope of -2.259, -2.533, and -3.240
for the star, ring, and 
squeezing ansatz, respectively. 
%Compared to the ring and squeezing ansatz, 
The star ansatz exhibits 
slower gradient decay as 
$\lambda$ approaches 1 
due to its smaller trainable variables' space
than the ring and squeezing ansatz.
%The slower decay of the star ansatz 
This indicates better training 
and less susceptibility 
to vanishing gradients, 
leading to better achievement 
of the ultimate quantum bound.

%\begin{widetext}
\begin{figure}[t]
\centering
\includegraphics[width=8.6cm]{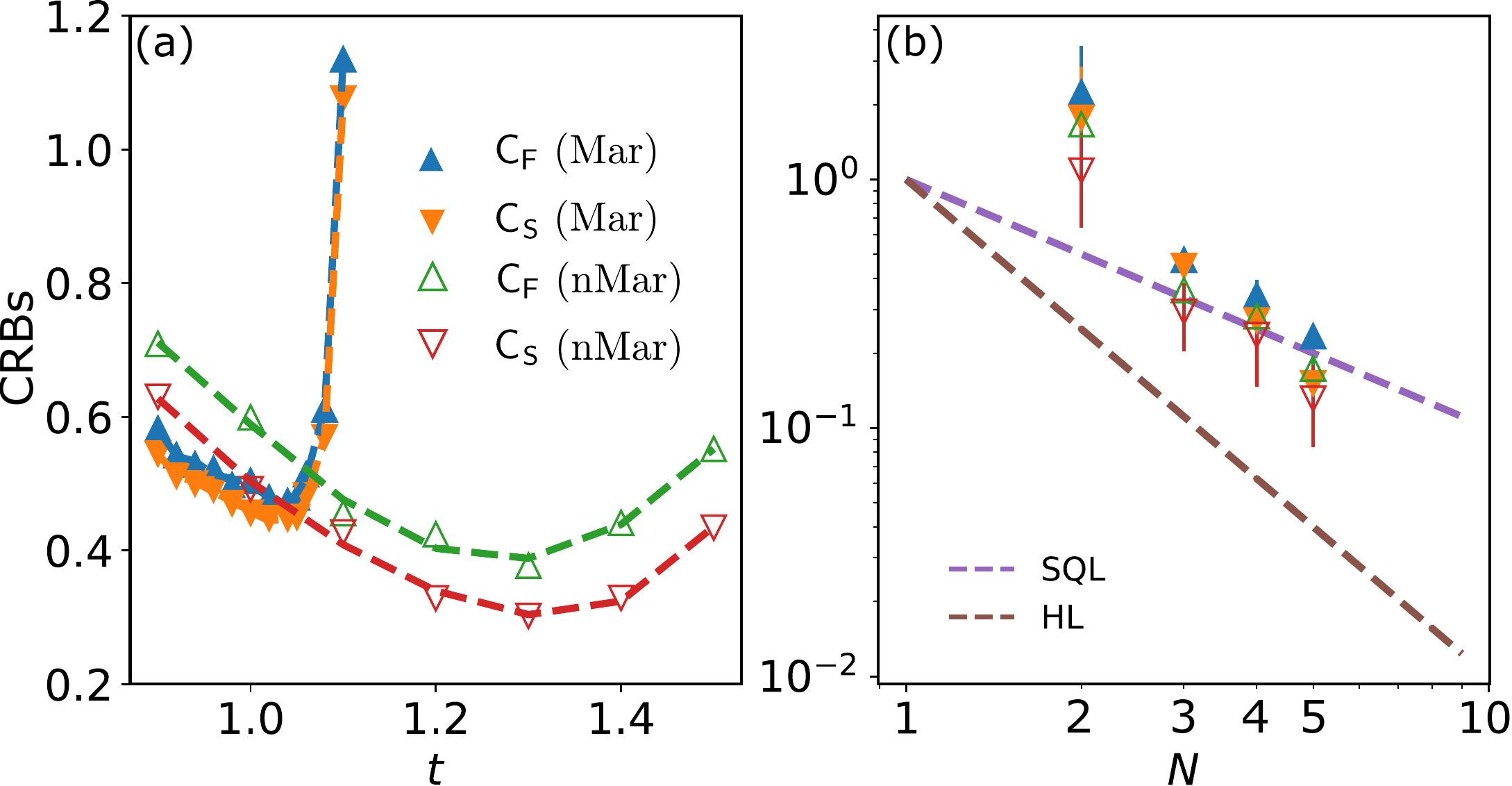}
\caption{
\textbf{Variational quantum metrology under time-dephasing noise}. 
(a): we present the CRBs as functions 
of the sensing time, demonstrating
an optimal sensing time for 
achieving each minimum CRB. 
The non-Markovian dephasing (nMar) 
produces lower metrological bounds 
in comparison to the Markovian one (Mar).
(b): plot of the minimal bounds 
for cases in (a), comparing them 
with the standard quantum limit (SQL) 
and the Heisenberg limit (HL). 
For non-Markovian metrology, 
the bounds surpass the SQL, 
as predicted.
}
\label{fig:5}
\end{figure} 
%\end{widetext}
%

\subsection{Multiparameter estimation under 
the Ornstein-Uhlenbeck model}

We consider the Ornstein-Uhlenbeck model, 
where the noise is induced by 
the stochastic fluctuation of 
the external (magnetic) field
\cite{PhysRev.36.823}. 
The Kraus operators are
\cite{YU2010676}
\begin{align}\label{eq:tdephasing}
K_1(t) = \begin{pmatrix} \sqrt{1-q(t)} & 0 \\ 0 & 1
\end{pmatrix},\
K_2(t) = \begin{pmatrix} \sqrt{q(t)} & 0 \\ 0 & 0
\end{pmatrix} ,
\end{align}
where $q(t) = 1-e^{-f(t)}$ 
with $f(t) = \gamma[t+\tau_c (e^{-t/\tau_c}-1)]$,
and $\tau_c$ represents 
the memory time of the environment.
In the Markovian limit ($\tau_c\to 0$), 
$f(t) = \gamma t$, which corresponds 
to the previous dephasing case.
In the non-Markovian limit 
with large $\tau_c$, such as $t/\tau_c\ll 1$,
we have $f(t) = \frac{\gamma t^2}{2\tau_c}$.
In the numerical simulation,
we fixed $\gamma = 0.1$
and $\tau_c = 20$ (for non-Markovian)
\begin{align}\label{eq:timefunc}
    q(t) = \begin{cases} 1-\exp(- 0.1t) \;\;\;\; &\textrm{Markovian,} \\ 
    1-\exp(-\frac{t^2}{400}) \;\;\; &\textrm{non-Markovian.} \end{cases}
\end{align}
We use this model to study 
the relationship between sensing time, 
Markovianity, 
and ultimate attainability 
of the quantum bound. 
Figure \ref{fig:5}a 
displays the optimal CRBs 
for Markovian and non-Markovian noises 
as functions of sensing time $t$.
As previously reported in 
\cite{PhysRevA.102.022602}, 
there exists an optimal 
sensing time that minimizes 
the CRBs for each case examined here.
%For Markovian (Mar) noise, 
%the optimal sensing time is shorter, 
%while for non-Markovian (nMar) noise, 
%it is longer.
Moreover, the non-Markovian dephasing (nMar)
results in lower metrological bounds 
as compared to the Markovian case (Mar).
So far, the minimum CRBs for different $N$ 
are presented in Figure \ref{fig:5}b. 
The results demonstrate that 
with an increase in $N$, 
the non-Markovian noise attains 
a better bound than the 
standard quantum limit (SQL) 
for both classical and SLD quantum bounds. 
This observation agrees with previous results 
reported using semidefinite programming 
\cite{PhysRevLett.127.060501}, 
indicating the potential of variational optimization 
for designing optimal non-Markovian metrology experiments.
Finally, we note that in 
the Ornstein-Uhlenbeck model, 
the SLD quantum bound is unachievable, 
as indicated by $\mathsf{C_F} > \mathsf{C_S}$. 
It remains a question for future research 
on whether one can attain the SLD quantum bound 
$\mathsf{C_S}$ with probe designs, 
and the existence of tight bounds 
in the non-Markovian scenario.

\section{Discussion}
\subsection{Concentratable entanglement}
We discuss how the three ansatzes 
create entangled states 
and the role of entangled resources 
in achieving the SLD quantum bound in VQM.
We analyze entanglement using 
the concentratable entanglement 
(CE) defined by 
\cite{PhysRevLett.127.140501}
\begin{align}\label{eq:CE}
    \xi(\psi) = 1 - \dfrac{1}{2^{|s|}} 
    \sum_{\alpha \in \mathcal{P}(s)} {\rm Tr}[\rho^2_\alpha],
\end{align}
where $\mathcal{P}(s)$
is the power set of $s,
\forall s\in \{1,2,\cdots,N\}$, 
and $\rho_\alpha$
is the reduced state of $|\psi\ra$ 
in the subsystem $\alpha$ 
with $\rho_\emptyset := \bm I$.
Practically, $\xi(\psi)$ can be computed 
using the SWAP test circuit as
stated in Ref. \cite{PhysRevLett.127.140501},
where
$
%\begin{align}\label{eq:appCE_sw}
    \xi(\psi) = 1 - p(\bm 0),
%\end{align}
$
with $p(\bm 0)$
is the probability of obtaining $|00\cdots 0\rangle$.
%(see Method for details.) 
%
The ability of the SWAP test to compute 
CE is due to the equivalence between 
conditional probability distribution 
and the definition of CE. 

%\subsection{Concentratable Entanglement}

\begin{figure}[t]
\centering
\includegraphics[width=8.6cm]{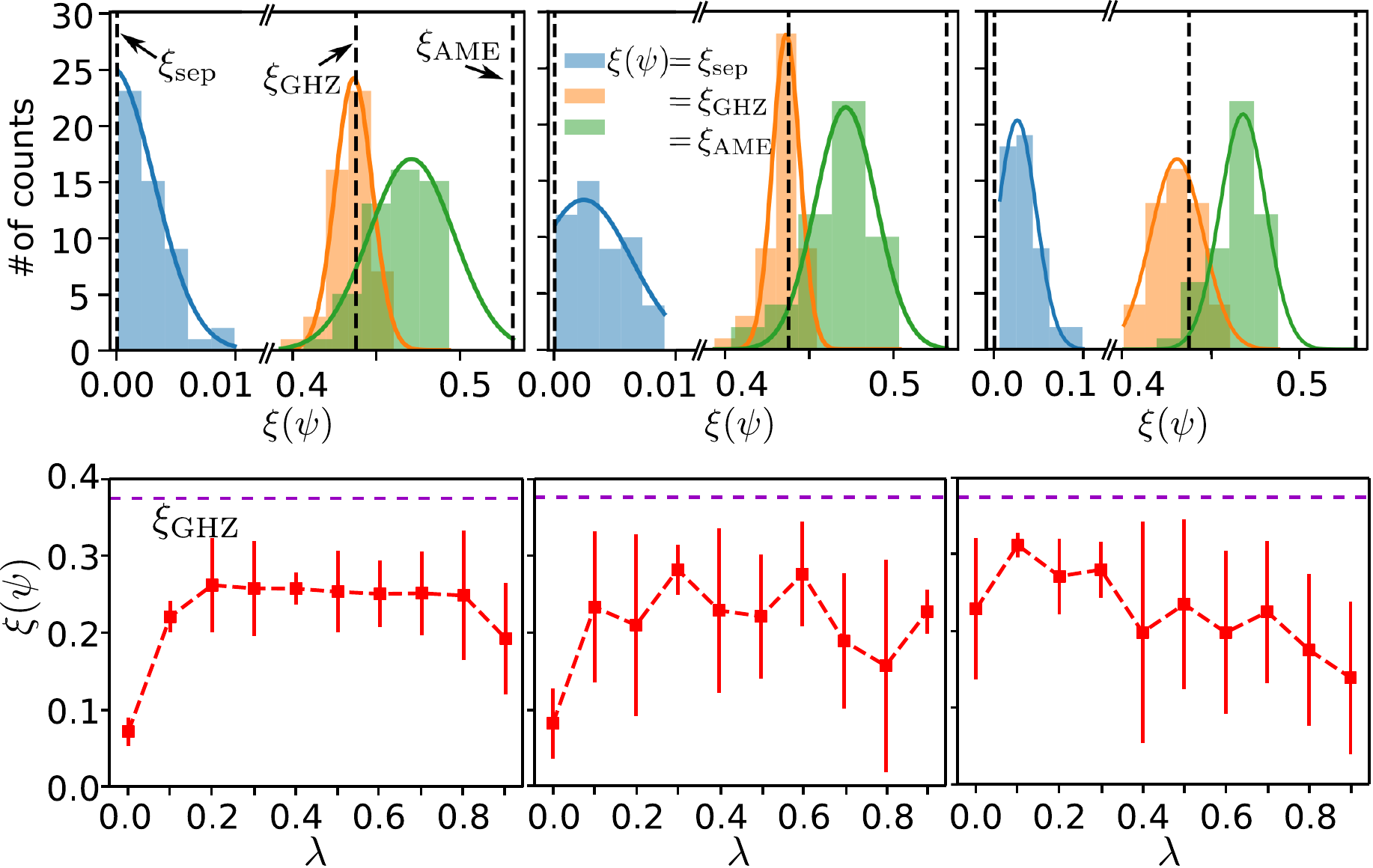}
\caption{
\textbf{Entanglement generation}. 
(Top): from left to right:
the distribution of training CEs 
corresponds to the star, ring, 
and squeezing ansatzes, 
respectively. 
All the ansatzes can produce 
separable and GHZ states, 
but generating an 
AME state is challenging. 
The results are shown at $N=4$
and (2-2) layers for each ansatz. 
(Bottom): the CEs are ploted at 
the optimal CRBs in Fig.~\ref{fig:3}, 
using the same circuit setup 
that in the figure. 
Again, $\lambda$ is the dephasing probability.
}
\label{fig:6}
\end{figure}

We first train the three ansatzes to evaluate
their ability of entangled-state generation.
Particularly, the training process aims to 
generate quantum states with
$\xi(\psi) =\{
\xi_{\rm sep}, \xi_{\rm GHZ}, \xi_{\rm AME}
\}$,
where $\xi_{\rm sep} = 0$ 
for a separable state, 
$\xi_{\rm GHZ} = 
\frac{1}{2} - \frac{1}{2^N}$ for a
GHZ state, 
and $\xi_{\rm AME} = 
1-\frac{1}{2^N}\sum_{j=0}^N
%{N \choose j}
\binom{N}{j}
\frac{1}{2^{\min(j, N-j)}}$
for an absolutely maximally entangled (AME) state
\cite{Enriquez_2016,PhysRevA.106.042411}.
The top panels in Fig.~\ref{fig:6} display
the results for star, ring,
and squeezing ansatz, 
from left to right, at $N = 4$ 
and (2-2) layers of each ansatz as an example.
All the ansatzes examined can reach 
the separable and GHZ state,
but hard to achieve the AME state.
This observation is consistent
with %the findings of 
the CEs for conventional graph states 
\cite{PhysRevA.106.042411}.

%which is equivalent to the GHZ state up to local unitaries

%\begin{figure}[t]
%\centering
%\includegraphics[width=5.6cm]{fig1_app.eps}
%\caption{
%\textbf{Swap test circuit}.
%A swap test circuit for computing CE.
%}
%\label{fig:1_app}
%\end{figure} 
%
We next discuss the role of entanglement 
in achieving the ultimate SLD quantum bound.
In the bottom panels of Fig.~\ref{fig:6}, 
we graph the corresponding CEs
at the optimal CRBs 
shown in Fig.~\ref{fig:3}, 
which apparently do not require 
the maximum entanglement (e.g., GHZ) 
to achieve the ultimate SLD quantum bound. 
This phenomenon can be explained by 
the fact that maximum entanglement 
is not required for high-precision 
quantum metrology, 
as previously noted in Refs.
\cite{PhysRevX.6.041044,RevModPhys.90.035006,PhysRevA.81.022108}. 
Therefore, emphasizing the robustness of 
easily preparable entangled probe states 
and non-local POVM schemes would be advantageous 
for quantum metrological applications exposed 
to Markovian and non-Markovian noises.

% maybe think more about bounds and non-Markovian quantum metrology in general 

% Wrapping up conclusion and future problems 
%\textcolor{red}{KL: To conclude, we have demonstrated }

\subsection{Cost functions}
\begin{figure}[t]
\centering
\includegraphics[width=8.6cm]{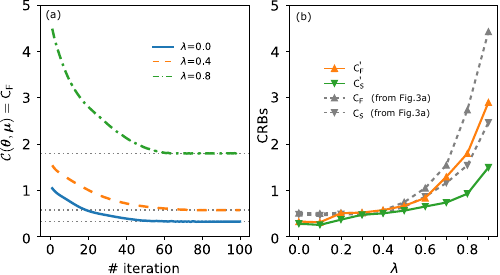}
\caption{
Optimizing classical bound $\mathsf{C_F}$ 
through Variational Quantum Metrology.
(a): Plot of the cost function $\mathcal{C}
(\bm\theta,\bm\mu) = \mathsf{C_F}$ 
versus the number of iterations 
across various noise probabilities $\lambda$.
(b) Plot of the corresponding 
optimal values of $\mathsf{C'_F}$ and 
$\mathsf{C'_S}$, and their counterparts 
extracted from Figure~\ref{fig:3}a (in grayscale).
These numerical results pertain 
to the star-graph configuration.
}
\label{fig:7}
\end{figure}

We address the selection of 
the cost function used in the variational algorithm.
The preference outlined in 
Eq.~\eqref{eq:cost} is not the sole option. 
An alternative approach could 
involve adopting the classical bound $\mathsf{C_F}$ 
as the cost function to maximize 
the information extraction.
However, this way does not guarantee 
the classical bound can reach the quantum bound,
%$\mathsf{C_F} = \mathsf{C_S}$, 
a requirement in estimation theory. 
In Figure~\ref{fig:7}a, we present a plot depicting the cost function $\mathcal{C}(\bm\theta,\bm\mu)=\mathsf{C_F}$ as a function of the number of iterations, considering various noise probabilities $\lambda$. It demonstrates that the cost function reaches its minimum value at a certain iteration.
Correspondingly, in Figure~\ref{fig:7}b, we provide the optimal values for both classical and quantum bounds, denoted as $\mathsf{C'_F}$ and $\mathsf{C'_S}$, alongside this optimization. It's important to emphasize that this approach does not guarantee that $\mathsf{C'_F}$ equals $\mathsf{C'_S}$.
For comparison, we include a grayscale representation of these quantities, originally presented in Figure~\ref{fig:3}a.
Here, optimizing the cost function Eq.\eqref{eq:cost} still ensures small values and convergence for both $\mathsf{C_F}$ and $\mathsf{C_S}$. However, $\mathsf{C'_F}$ and $\mathsf{C'_S}$ consistently remain below $\mathsf{C_F}$ and $\mathsf{C_S}$. This behavior occurs because the evaluation of $\mathsf{C_F}$ and $\mathsf{C_S}$ is based on the state that maximizes the figure of merit in Eq.~\eqref{eq:cost}, rather than solely minimizing $\mathsf{C_F}$ or $\mathsf{C_S}$.

\begin{figure*}[t]
\centering
\includegraphics[width=15.50cm]{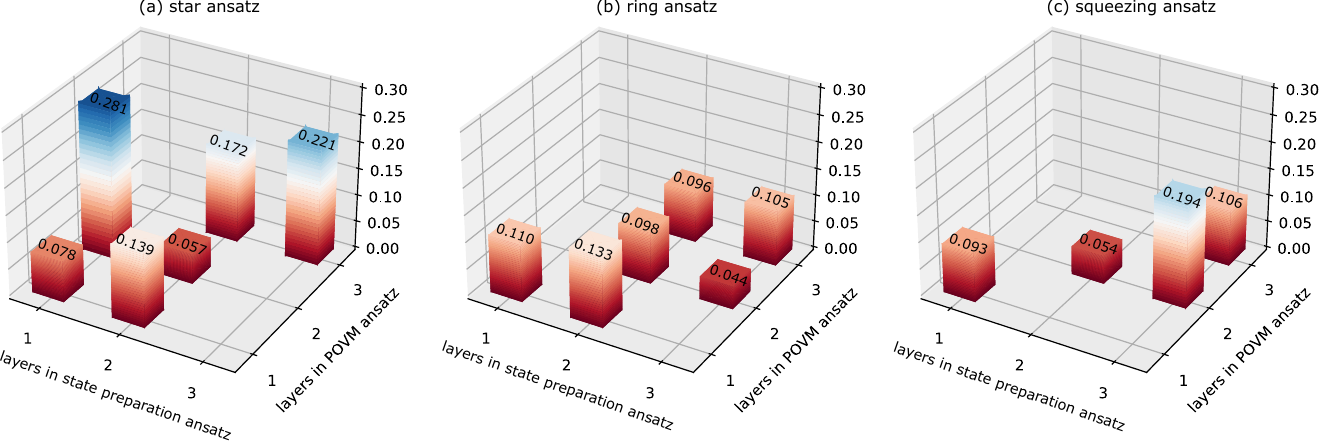}
\caption{
\textbf{Cost function vs the optimal number 
of layers for different ansatzes}.
The plot of the cost function for
(a) star ansatz, (b) ring ansatz,
and (c) squeezing ansatz at $N = 3$ without noise.
The minimum cost function vs the number of layers are
(0.057, 2-2), (0.044, 3-2), and (0.054, 2-2), 
respectively.
}
\label{fig:8}
\end{figure*} 

Furthermore, alternative physical quantities, 
such as the tradeoff $\mathcal{T}$ 
and the norm-2, can also be utilized 
as potential cost functions. 
For instance, a tradeoff 
cost function is analogous
to the one presented in 
Eq.~\eqref{eq:cost}, taking the form: 
\begin{align}\label{eq:trof} 
\mathcal{C}(\bm\theta,\bm\mu) = 
1 - \dfrac{1}{d} {\rm Tr}[FQ^{-1}], 
\end{align} 
where $d$ is
the number of estimated parameters. 
Notably, in scenarios where $Q$ is
a diagonal matrix and 
$\mathcal{C}(\bm\theta,\bm\mu) = 0$, 
it results in a zero tradeoff, i.e., 
$\frac{F_{11}}{Q_{11}} =\cdots = 
\frac{F_{dd}}{Q_{dd}} = 1$.
Additionally, norm-2 can also function 
as a viable cost function \cite{PhysRevLett.119.130504}. 
\begin{align}\label{eq:norm2} 
\mathcal{C}(\bm\theta,\bm\mu) 
= ||F-Q||_2, 
\end{align} 
where $||A||_2 = \sqrt{\lambda_{\max}(A^*A)}$
represents the norm-2,
with $\lambda_{\max}$ is the maximum eigenvalue.
However, it is worth noting that these alternate cost functions 
might not be convex nor trainable \cite{Cerezo2021}. 
As a result, the selection of 
the cost function in  Eq.~\eqref{eq:cost} 
is indeed appropriate.

\section{Methods}
\subsection{Quantum circuit training} 
In numerical simulations, 
we employ the ADAM optimizer 
to train the VQM variables \cite{kingma2015adam},
where the variables at step $k+1$ are given by
\begin{align}\label{eq:adam}
\bm{\theta}^{k+1}=\bm{\theta}^{k}
-\alpha\frac{\hat{m}_{k}}{\sqrt{\hat{v}_{k}} + \epsilon},
\end{align}
where $m_{k}=\beta_{1} m_{k-1}
+\left(1-\beta_{1}\right) 
\nabla_{\bm\theta}\mathcal{C}(\bm\theta), 
v_{k}=\beta_{2} v_{k-1}+(1-\beta_{2}) 
\nabla_{\bm\theta}^2\mathcal{C}(\bm\theta),
\hat{m}_{k}=m_{k} /\left(1-\beta_{1}^{k}\right),
\hat{v}_{k}=v_{k} /\left(1-\beta_{2}^{k}\right),
$
with the hyper-parameters 
are chosen as 
$\alpha = 0.2, \beta_1 = 0.8, 
\beta_2 = 0.999$ 
and $\epsilon = 10^{-8}$. 
The gradient $\partial_{\theta_i}\mathcal{C}(\bm\theta)$
is given through the parameter-shift rule
\cite{PhysRevA.98.032309,PhysRevA.99.032331}.
The simulations are performed 
in Qiskit Aer simulator 
\cite{qiskit_aer_provider_tutorial}. 
%We applied the paramater-shift rule 
%to compute derivative w.r.t. training parameters, 
%and finite difference method 
%to compute the Fisher information.
The number of iterations is 
chosen using the EarlyStopping callback
\cite{keras_early_stopping}. 

To determine the appropriate number of layers 
for the preparation state and POVM ansatzes, 
we analyze the cost function \eqref{eq:cost} 
with different number of layers. 
We use $(\star,\dagger$-$\ddagger)$ to denote 
the minimum cost function, the number of layers 
for variational state preparation, and the number 
of layers for variational POVM.
The results are shown in Fig.~\ref{fig:8}
with $(\star,\dagger$-$\ddagger)=$ 
(0.057, 2-2), (0.04, 3-2), and (0.054, 2-2)
for the star, ring, and squeezing ansatz,
respectively. 
Obviously, the metrological performances 
of these ansatzes demonstrate that deep 
ansatzes are unnecessary, as also noted in 
\cite{https://doi.org/10.48550/arxiv.2008.01502} where 
a shallow ansatz was able to saturate quantum bound.
For the numerical simulations presented in this paper, 
we keep the number of layers fixed at these values.

%\subsection{Fisher information matrices}
\subsection{Computing Fisher information} 
Classical and quantum Fisher 
information matrices
can be computed in quantum circuits 
using the %parameter-shift rule. 
finite difference approximation. 
For the CFIM, we first derive 
an output probability as
%\begin{align}\label{eq:dp}
$
    \partial_{\phi_i} p
    = \frac{p(\phi_i + s) - p(\phi_i - s)}
    {2s},
$    
%\end{align}
for a small shift $s$.
We then compute the CFIM from
$F_{ij} 
= \sum_m 
\frac{1}{p(m|\bm\phi)}
[\partial_{\phi_i} p(m|\bm\phi)]
[\partial_{\phi_j} p(m|\bm \phi)]$.
For the QFIM, we explicitly derive
%\begin{align}\label{eq:sldqfimvec}
$
    Q_{ij} = 2{\rm vec}
    [\partial_{\phi_i}\rho(\bm\phi)]^\dagger
    \big[{\rho(\bm\phi)}^*\otimes 
    \bm I+\bm I\otimes \rho(\bm\phi)\big]^+
    {\rm vec}[\partial_{\phi_j}\rho(\bm\phi)],
$    
%\end{align}
where ${\rm vec}[\cdot]$
is the vectorization of a matrix,
and the superscript `+' 
denotes the 
pseudo-inversion \cite{PhysRevA.97.042322}.
%Similarly, the RLD QFIM yields
%\begin{align}\label{eq:rldqfimvec}
%    Q_{ij}^\mathsf{R} = 
%    \big[(\bm I\otimes \rho)^+
%    {\rm vec}[\partial_j\rho]
%    \big]^\dagger
%    {\rm vec}[\partial_i\rho].
%\end{align}
Again, we apply the 
finite difference to compute
%\begin{align}\label{eq:drho} 
$
    \partial_{\phi_i} \rho
    = \frac{\rho(\phi_i + s) 
    - \rho(\phi_i - s)}{2s},
$    
%\end{align}
and substitute into 
the above equations 
to compute the QFIM.

%\section*{Code availability}
All codes used to produce the findings of this study are 
incorporated into \texttt{tqix} \cite{VIET2023108686,HO2021107902}
and available at: https://github.com/echkon/tqix-developers.
See also the Supplementary Material for tutorial codes.

\begin{acknowledgments}
We thank C.Q. Nguyen for assisting with the 
initial code. This work is supported by 
JSPS KAKENHI Grant Number 23K13025.
\end{acknowledgments}

\bibliographystyle{unsrt}
\bibliography{refs}

\begin{thebibliography}{10}

\bibitem{PhysRevLett.72.3439}
Samuel~L. Braunstein and Carlton~M. Caves.
\newblock Statistical distance and the geometry of quantum states.
\newblock {\em Phys. Rev. Lett.}, 72:3439--3443, May 1994.

\bibitem{Giovannetti2011}
Vittorio Giovannetti, Seth Lloyd, and Lorenzo Maccone.
\newblock Advances in quantum metrology.
\newblock {\em Nature Photonics}, 5(4):222--229, Apr 2011.

\bibitem{Toth_2014}
Géza Tóth and Iagoba Apellaniz.
\newblock Quantum metrology from a quantum information science perspective.
\newblock {\em Journal of Physics A: Mathematical and Theoretical},
  47(42):424006, oct 2014.

\bibitem{Barndorff-Nielsen_2000}
O~E Barndorff-Nielsen and R~D Gill.
\newblock Fisher information in quantum statistics.
\newblock {\em Journal of Physics A: Mathematical and General}, 33(24):4481,
  jun 2000.

\bibitem{Fujiwara_2006}
Akio Fujiwara.
\newblock Strong consistency and asymptotic efficiency for adaptive quantum
  estimation problems.
\newblock {\em Journal of Physics A: Mathematical and General}, 39(40):12489,
  sep 2006.

\bibitem{Zhang_2018}
Yi-Hao Zhang and Wen Yang.
\newblock Improving spin-based noise sensing by adaptive measurements.
\newblock {\em New Journal of Physics}, 20(9):093011, sep 2018.

\bibitem{PhysRevX.7.041009}
Rafa\l{} Demkowicz-Dobrza\ifmmode~\acute{n}\else \'{n}\fi{}ski, Jan Czajkowski,
  and Pavel Sekatski.
\newblock Adaptive quantum metrology under general markovian noise.
\newblock {\em Phys. Rev. X}, 7:041009, Oct 2017.

\bibitem{PhysRevLett.112.150802}
E.~M. Kessler, I.~Lovchinsky, A.~O. Sushkov, and M.~D. Lukin.
\newblock Quantum error correction for metrology.
\newblock {\em Phys. Rev. Lett.}, 112:150802, Apr 2014.

\bibitem{Zhou2018}
Sisi Zhou, Mengzhen Zhang, John Preskill, and Liang Jiang.
\newblock Achieving the heisenberg limit in quantum metrology using quantum
  error correction.
\newblock {\em Nature Communications}, 9(1):78, Jan 2018.

\bibitem{PhysRevLett.128.160505}
Jing Yang, Shengshi Pang, Zekai Chen, Andrew~N. Jordan, and Adolfo del Campo.
\newblock Variational principle for optimal quantum controls in quantum
  metrology.
\newblock {\em Phys. Rev. Lett.}, 128:160505, Apr 2022.

\bibitem{Pang2017}
Shengshi Pang and Andrew~N. Jordan.
\newblock Optimal adaptive control for quantum metrology with time-dependent
  hamiltonians.
\newblock {\em Nature Communications}, 8(1):14695, Mar 2017.

\bibitem{Yang2021}
Xiaodong Yang, Xi~Chen, Jun Li, Xinhua Peng, and Raymond Laflamme.
\newblock Hybrid quantum-classical approach to enhanced quantum metrology.
\newblock {\em Scientific Reports}, 11(1):672, Jan 2021.

\bibitem{Koczor_2020}
Bálint Koczor, Suguru Endo, Tyson Jones, Yuichiro Matsuzaki, and Simon~C
  Benjamin.
\newblock Variational-state quantum metrology.
\newblock {\em New Journal of Physics}, 22(8):083038, aug 2020.

\bibitem{9605341}
Ziqi Ma, Pranav Gokhale, Tian-Xing Zheng, Sisi Zhou, Xiaofei Yu, Liang Jiang,
  Peter Maurer, and Frederic~T. Chong.
\newblock Adaptive circuit learning for quantum metrology.
\newblock In {\em 2021 IEEE International Conference on Quantum Computing and
  Engineering (QCE)}, pages 419--430, 2021.

\bibitem{PhysRevLett.123.260505}
Raphael Kaubruegger, Pietro Silvi, Christian Kokail, Rick van Bijnen, Ana~Maria
  Rey, Jun Ye, Adam~M. Kaufman, and Peter Zoller.
\newblock Variational spin-squeezing algorithms on programmable quantum
  sensors.
\newblock {\em Phys. Rev. Lett.}, 123:260505, Dec 2019.

\bibitem{PhysRevX.11.041045}
Raphael Kaubruegger, Denis~V. Vasilyev, Marius Schulte, Klemens Hammerer, and
  Peter Zoller.
\newblock Quantum variational optimization of ramsey interferometry and atomic
  clocks.
\newblock {\em Phys. Rev. X}, 11:041045, Dec 2021.

\bibitem{Zheng2022}
Tian-Xing Zheng, Anran Li, Jude Rosen, Sisi Zhou, Martin Koppenh{\"o}fer, Ziqi
  Ma, Frederic~T. Chong, Aashish~A. Clerk, Liang Jiang, and Peter~C. Maurer.
\newblock Preparation of metrological states in dipolar-interacting spin
  systems.
\newblock {\em npj Quantum Information}, 8(1):150, Dec 2022.

\bibitem{PhysRevLett.124.160502}
Zhi Li, Liujun Zou, and Timothy~H. Hsieh.
\newblock Hamiltonian tomography via quantum quench.
\newblock {\em Phys. Rev. Lett.}, 124:160502, Apr 2020.

\bibitem{PhysRevLett.116.030801}
Tillmann Baumgratz and Animesh Datta.
\newblock Quantum enhanced estimation of a multidimensional field.
\newblock {\em Phys. Rev. Lett.}, 116:030801, Jan 2016.

\bibitem{PhysRevA.102.022602}
Le~Bin Ho, Hideaki Hakoshima, Yuichiro Matsuzaki, Masayuki Matsuzaki, and
  Yasushi Kondo.
\newblock Multiparameter quantum estimation under dephasing noise.
\newblock {\em Phys. Rev. A}, 102:022602, Aug 2020.

\bibitem{PhysRevLett.125.020501}
Zhibo Hou, Zhao Zhang, Guo-Yong Xiang, Chuan-Feng Li, Guang-Can Guo, Hongzhen
  Chen, Liqiang Liu, and Haidong Yuan.
\newblock Minimal tradeoff and ultimate precision limit of multiparameter
  quantum magnetometry under the parallel scheme.
\newblock {\em Phys. Rev. Lett.}, 125:020501, Jul 2020.

\bibitem{Schnabel2010}
Roman Schnabel, Nergis Mavalvala, David~E. McClelland, and Ping~K. Lam.
\newblock Quantum metrology for gravitational wave astronomy.
\newblock {\em Nature Communications}, 1(1):121, Nov 2010.

\bibitem{RevModPhys.83.331}
Andrei Derevianko and Hidetoshi Katori.
\newblock Colloquium: Physics of optical lattice clocks.
\newblock {\em Rev. Mod. Phys.}, 83:331--347, May 2011.

\bibitem{RevModPhys.87.637}
Andrew~D. Ludlow, Martin~M. Boyd, Jun Ye, E.~Peik, and P.~O. Schmidt.
\newblock Optical atomic clocks.
\newblock {\em Rev. Mod. Phys.}, 87:637--701, Jun 2015.

\bibitem{PhysRevX.12.011039}
Francesco Albarelli and Rafa\l{} Demkowicz-Dobrza\ifmmode~\acute{n}\else
  \'{n}\fi{}ski.
\newblock Probe incompatibility in multiparameter noisy quantum metrology.
\newblock {\em Phys. Rev. X}, 12:011039, Mar 2022.

\bibitem{Kull_2020}
Ilya Kull, Philippe~Allard Guérin, and Frank Verstraete.
\newblock Uncertainty and trade-offs in quantum multiparameter estimation.
\newblock {\em Journal of Physics A: Mathematical and Theoretical},
  53(24):244001, may 2020.

\bibitem{PhysRevLett.119.130504}
Luca Pezz\`e, Mario~A. Ciampini, Nicol\`o Spagnolo, Peter~C. Humphreys, Animesh
  Datta, Ian~A. Walmsley, Marco Barbieri, Fabio Sciarrino, and Augusto Smerzi.
\newblock Optimal measurements for simultaneous quantum estimation of multiple
  phases.
\newblock {\em Phys. Rev. Lett.}, 119:130504, Sep 2017.

\bibitem{PhysRevA.100.032104}
Jing Yang, Shengshi Pang, Yiyu Zhou, and Andrew~N. Jordan.
\newblock Optimal measurements for quantum multiparameter estimation with
  general states.
\newblock {\em Phys. Rev. A}, 100:032104, Sep 2019.

\bibitem{PhysRevLett.117.160801}
Haidong Yuan.
\newblock Sequential feedback scheme outperforms the parallel scheme for
  hamiltonian parameter estimation.
\newblock {\em Phys. Rev. Lett.}, 117:160801, Oct 2016.

\bibitem{10.1063/5.0024555}
Le~Bin Ho and Yasushi Kondo.
\newblock {Multiparameter quantum metrology with postselection measurements}.
\newblock {\em Journal of Mathematical Physics}, 62(1):012102, 01 2021.

\bibitem{Meyer2021}
Johannes~Jakob Meyer, Johannes Borregaard, and Jens Eisert.
\newblock A variational toolbox for quantum multi-parameter estimation.
\newblock {\em npj Quantum Information}, 7(1):89, Jun 2021.

\bibitem{Zhu2015}
Huangjun Zhu.
\newblock Information complementarity: A new paradigm for decoding quantum
  incompatibility.
\newblock {\em Scientific Reports}, 5(1):14317, Sep 2015.

\bibitem{PhysRevA.94.052108}
Sammy Ragy, Marcin Jarzyna, and Rafa\l{}
  Demkowicz-Dobrza\ifmmode~\acute{n}\else \'{n}\fi{}ski.
\newblock Compatibility in multiparameter quantum metrology.
\newblock {\em Phys. Rev. A}, 94:052108, Nov 2016.

\bibitem{PhysRev.36.823}
G.~E. Uhlenbeck and L.~S. Ornstein.
\newblock On the theory of the brownian motion.
\newblock {\em Phys. Rev.}, 36:823--841, Sep 1930.

\bibitem{helstrom1976quantum}
C.~W. Helstrom.
\newblock {\em Quantum Detection and Estimation Theory}.
\newblock Academic Press, New York, 1976.

\bibitem{holevo2011probabilistic}
A.S. Holevo.
\newblock {\em Probabilistic and Statistical Aspects of Quantum Theory}.
\newblock Springer, New York, 1st edition, 2011.

\bibitem{conlon2022gap}
Lorcán~O. Conlon, Jun Suzuki, Ping~Koy Lam, and Syed~M. Assad.
\newblock The gap persistence theorem for quantum multiparameter estimation,
  2022.

\bibitem{hayashi2023tight}
Masahito Hayashi and Yingkai Ouyang.
\newblock Tight cram\'{e}r-rao type bounds for multiparameter quantum metrology
  through conic programming, 2023.

\bibitem{paris2009quantum}
Marco G~A Paris.
\newblock Quantum state estimation.
\newblock {\em International Journal of Quantum Information}, 7(1):125--137,
  2009.

\bibitem{Conlon2021}
Lorc{\'a}n~O. Conlon, Jun Suzuki, Ping~Koy Lam, and Syed~M. Assad.
\newblock Efficient computation of the nagaoka--hayashi bound for
  multiparameter estimation with separable measurements.
\newblock {\em npj Quantum Information}, 7(1):110, Jul 2021.

\bibitem{hayashi2005}
M.~Hayashi, editor.
\newblock {\em Asymptotic Theory of Quantum Statistical Inference: Selected
  Papers}.
\newblock World Scientific Singapore, 2005.

\bibitem{kay1993estimation}
Steven Kay.
\newblock {\em Estimation theory, Vol 1}.
\newblock Prentice Hall, Englewood Cliffs, NJ, 1st edition, 1993.

\bibitem{Yang2019}
Yuxiang Yang, Giulio Chiribella, and Masahito Hayashi.
\newblock Attaining the ultimate precision limit in quantum state estimation.
\newblock {\em Communications in Mathematical Physics}, 368(1):223--293, May
  2019.

\bibitem{10.1214/13-AOS1147}
Koichi Yamagata, Akio Fujiwara, and Richard~D. Gill.
\newblock {Quantum local asymptotic normality based on a new quantum likelihood
  ratio}.
\newblock {\em The Annals of Statistics}, 41(4):2197 -- 2217, 2013.

\bibitem{PhysRevX.11.011028}
Jasminder~S. Sidhu, Yingkai Ouyang, Earl~T. Campbell, and Pieter Kok.
\newblock Tight bounds on the simultaneous estimation of incompatible
  parameters.
\newblock {\em Phys. Rev. X}, 11:011028, Feb 2021.

\bibitem{https://doi.org/10.48550/arxiv.2008.01502}
Jamie Friel, Pantita Palittapongarnpim, Francesco Albarelli, and Animesh Datta.
\newblock Attainability of the holevo-cramér-rao bound for two-qubit 3d
  magnetometry, 2020.

\bibitem{PhysRevLett.123.200503}
Francesco Albarelli, Jamie~F. Friel, and Animesh Datta.
\newblock Evaluating the holevo cram\'er-rao bound for multiparameter quantum
  metrology.
\newblock {\em Phys. Rev. Lett.}, 123:200503, Nov 2019.

\bibitem{Demkowicz-Dobrzanski_2020}
Rafał Demkowicz-Dobrzański, Wojciech Górecki, and Mădălin Guţă.
\newblock Multi-parameter estimation beyond quantum fisher information.
\newblock {\em Journal of Physics A: Mathematical and Theoretical},
  53(36):363001, aug 2020.

\bibitem{K_Matsumoto_2002}
K~Matsumoto.
\newblock A new approach to the cramér-rao-type bound of the pure-state model.
\newblock {\em Journal of Physics A: Mathematical and General}, 35(13):3111,
  mar 2002.

\bibitem{PhysRevLett.111.070403}
Peter~C. Humphreys, Marco Barbieri, Animesh Datta, and Ian~A. Walmsley.
\newblock Quantum enhanced multiple phase estimation.
\newblock {\em Phys. Rev. Lett.}, 111:070403, Aug 2013.

\bibitem{PhysRevLett.124.110502}
Nathan Shettell and Damian Markham.
\newblock Graph states as a resource for quantum metrology.
\newblock {\em Phys. Rev. Lett.}, 124:110502, Mar 2020.

\bibitem{PhysRevA.102.052601}
Yunkai Wang and Kejie Fang.
\newblock Continuous-variable graph states for quantum metrology.
\newblock {\em Phys. Rev. A}, 102:052601, Nov 2020.

\bibitem{Maccone2020squeezingmetrology}
Lorenzo Maccone and Alberto Riccardi.
\newblock Squeezing metrology: a unified framework.
\newblock {\em {Quantum}}, 4:292, July 2020.

\bibitem{Gessner2020}
Manuel Gessner, Augusto Smerzi, and Luca Pezz{\`e}.
\newblock Multiparameter squeezing for optimal quantum enhancements in sensor
  networks.
\newblock {\em Nature Communications}, 11(1):3817, Jul 2020.

\bibitem{PhysRevA.102.062610}
G.~Carrara, M.~G. Genoni, S.~Cialdi, M.~G.~A. Paris, and S.~Olivares.
\newblock Squeezing as a resource to counteract phase diffusion in optical
  phase estimation.
\newblock {\em Phys. Rev. A}, 102:062610, Dec 2020.

\bibitem{PhysRevLett.82.1835}
Klaus M\o{}lmer and Anders S\o{}rensen.
\newblock Multiparticle entanglement of hot trapped ions.
\newblock {\em Phys. Rev. Lett.}, 82:1835--1838, Mar 1999.

\bibitem{Koczor2020}
Bálint Koczor, Suguru Endo, Tyson Jones, Yuichiro Matsuzaki, and Simon~C
  Benjamin.
\newblock Variational-state quantum metrology.
\newblock {\em New Journal of Physics}, 22(8):083038, aug 2020.

\bibitem{keras_early_stopping}
Keras.
\newblock Earlystopping callback - keras documentation.
\newblock \url{https://keras.io/api/callbacks/early_stopping/}, 2021.
\newblock [Online; accessed 28-March-2023].

\bibitem{Vidrighin2014}
Mihai~D. Vidrighin, Gaia Donati, Marco~G. Genoni, Xian-Min Jin, W.~Steven
  Kolthammer, M.~S. Kim, Animesh Datta, Marco Barbieri, and Ian~A. Walmsley.
\newblock Joint estimation of phase and phase diffusion for quantum metrology.
\newblock {\em Nature Communications}, 5(1):3532, Apr 2014.

\bibitem{Wang2021}
Samson Wang, Enrico Fontana, M.~Cerezo, Kunal Sharma, Akira Sone, Lukasz
  Cincio, and Patrick~J. Coles.
\newblock Noise-induced barren plateaus in variational quantum algorithms.
\newblock {\em Nature Communications}, 12(1):6961, Nov 2021.

\bibitem{Cerezo2021}
M.~Cerezo, Andrew Arrasmith, Ryan Babbush, Simon~C. Benjamin, Suguru Endo,
  Keisuke Fujii, Jarrod~R. McClean, Kosuke Mitarai, Xiao Yuan, Lukasz Cincio,
  and Patrick~J. Coles.
\newblock Variational quantum algorithms.
\newblock {\em Nature Reviews Physics}, 3(9):625--644, Sep 2021.

\bibitem{YU2010676}
Ting Yu and J.H. Eberly.
\newblock Entanglement evolution in a non-markovian environment.
\newblock {\em Optics Communications}, 283(5):676--680, 2010.
\newblock Quo vadis Quantum Optics?

\bibitem{PhysRevLett.127.060501}
Anian Altherr and Yuxiang Yang.
\newblock Quantum metrology for non-markovian processes.
\newblock {\em Phys. Rev. Lett.}, 127:060501, Aug 2021.

\bibitem{PhysRevLett.127.140501}
Jacob~L. Beckey, N.~Gigena, Patrick~J. Coles, and M.~Cerezo.
\newblock Computable and operationally meaningful multipartite entanglement
  measures.
\newblock {\em Phys. Rev. Lett.}, 127:140501, Sep 2021.

\bibitem{Enriquez_2016}
M~Enríquez, I~Wintrowicz, and K~Życzkowski.
\newblock Maximally entangled multipartite states: A brief survey.
\newblock {\em Journal of Physics: Conference Series}, 698(1):012003, mar 2016.

\bibitem{PhysRevA.106.042411}
Alice~R. Cullen and Pieter Kok.
\newblock Calculating concentratable entanglement in graph states.
\newblock {\em Phys. Rev. A}, 106:042411, Oct 2022.

\bibitem{PhysRevX.6.041044}
M.~Oszmaniec, R.~Augusiak, C.~Gogolin, J.~Ko\l{}ody\ifmmode~\acute{n}\else
  \'{n}\fi{}ski, A.~Ac\'{\i}n, and M.~Lewenstein.
\newblock Random bosonic states for robust quantum metrology.
\newblock {\em Phys. Rev. X}, 6:041044, Dec 2016.

\bibitem{RevModPhys.90.035006}
Daniel Braun, Gerardo Adesso, Fabio Benatti, Roberto Floreanini, Ugo Marzolino,
  Morgan~W. Mitchell, and Stefano Pirandola.
\newblock Quantum-enhanced measurements without entanglement.
\newblock {\em Rev. Mod. Phys.}, 90:035006, Sep 2018.

\bibitem{PhysRevA.81.022108}
Todd Tilma, Shinichiro Hamaji, W.~J. Munro, and Kae Nemoto.
\newblock Entanglement is not a critical resource for quantum metrology.
\newblock {\em Phys. Rev. A}, 81:022108, Feb 2010.

\bibitem{kingma2015adam}
Diederik~P Kingma and Jimmy Ba.
\newblock Adam: A method for stochastic optimization.
\newblock In {\em Proceedings of the 3rd International Conference on Learning
  Representations (ICLR)}, 2015.

\bibitem{PhysRevA.98.032309}
K.~Mitarai, M.~Negoro, M.~Kitagawa, and K.~Fujii.
\newblock Quantum circuit learning.
\newblock {\em Phys. Rev. A}, 98:032309, Sep 2018.

\bibitem{PhysRevA.99.032331}
Maria Schuld, Ville Bergholm, Christian Gogolin, Josh Izaac, and Nathan
  Killoran.
\newblock Evaluating analytic gradients on quantum hardware.
\newblock {\em Phys. Rev. A}, 99:032331, Mar 2019.

\bibitem{qiskit_aer_provider_tutorial}
Qiskit.
\newblock Aer provider tutorial, 2021.

\bibitem{PhysRevA.97.042322}
Dominik \ifmmode~\check{S}\else \v{S}\fi{}afr\'anek.
\newblock Simple expression for the quantum fisher information matrix.
\newblock {\em Phys. Rev. A}, 97:042322, Apr 2018.

\bibitem{VIET2023108686}
Nguyen~Tan Viet, Nguyen~Thi Chuong, Vu~Thi~Ngoc Huyen, and Le~Bin Ho.
\newblock tqix.pis: A toolbox for quantum dynamics simulation of spin ensembles
  in dicke basis.
\newblock {\em Computer Physics Communications}, 286:108686, 2023.

\bibitem{HO2021107902}
Le~Bin Ho, Kieu~Quang Tuan, and Hung~Q. Nguyen.
\newblock tqix: A toolbox for quantum in x: X: Quantum measurement, quantum
  tomography, quantum metrology, and others.
\newblock {\em Computer Physics Communications}, 263:107902, 2021.

\end{thebibliography}

\end{document}